\documentclass[useAMS,usenatbib]{mn2e}
\usepackage{graphicx,graphics,amsmath}

\newcommand{\Om}{\Omega_{\rm M}}

\newcommand{\Ox}{\Omega_{\rm X}}

\begin{document}
\title {High-z Supernova Type Ia Data: non-Gaussianity and Direction Dependence}

\author[Shashikant Gupta and Meghendra Singh]
{Shashikant Gupta $^{1,2}$ and Meghendra Singh $^{3,4}$\\
  $^{1}$ Amity University Haryana, Gurgaon, India \\
  $^{2}$ shashikantgupta.astro@gmail.com \\
  $^{3}$ Mahamaya Tech. University, Noida, India \\
  $^{4}$ meghendrasingh\_db@yahoo.co.in \\
} \maketitle

\begin{abstract}
  We use the $\Delta_{\chi^2}$ statistic introduced in \cite{gup08,gup10}
  to study directional dependence, in the high-z supernovae data. 
  This dependence could arise due to departures from 
  the cosmological principle or from direction dependent statistical systematics in
  the data.  We apply our statistic to the 
  gold data set from \cite{rie04} and \cite{rie07}, and Union2 catalogue from \cite{aman10}.
  Our results show that all the three data sets show a weak but consistent direction 
  dependence. In 2007 data errors are Gaussian, however other two data sets show 
  non-Gaussian features.
\end{abstract}

\section{Introduction}
%***********************
A large and diverse variety of cosmological observations 
\citep{per99,rie98,rie02,rie04,rie07,ben02,pag06}, during the last 
two decades (notably the Nobel prizes in 2006 and in 2011) have established 
that we live in a flat universe with an accelerated expansion. This is consistent
with the Einstein's general theory of relativity with a cosmological 
constant term. The Cosmological Constant ($\Lambda$), also known as dark energy, 
is treated as an ideal fluid with negative pressure; and with equation of state  
$p = w\rho$, where $w=-1$. Dark energy density dominates over that of baryonic and 
dark matter; and it constitute two third of the Universe. This model of the Universe 
is known as the $\Lambda$CDM model or standard model of cosmology.

The foundation of the standard model of cosmology ($\Lambda$CDM cosmology) is
the Cosmological principle (hereafter CP), which states that the Universe is 
homogeneous and isotropic on the large scales \citep{pee93}. CP along with the fact 
that two third of the constituents of the Universe is dark energy can explain many 
cosmological observations and is the most successful model till date. Despite its success, 
$\Lambda$CDM model has its shortcomings; few of which are summarized below: 
\begin{itemize}
\item Observations indicate significantly larger amplitude of 
flows than what $\Lambda$CDM predicts. \citep{wat08} found the 
large scale peculiar velocity larger than 400km/Sec at scales up to 
$100 h^{-1}$ Mpc. This is in the direction $l=282^{\circ}$, $b=6^{\circ}$. 
The probability of finding such a flow in the $\Lambda$CDM cosmology 
is less than $1\%$.
\item The planes normal to the quadrupole ($l=240^{\circ}, b= 63^{\circ}$) and 
octopole ($l=308^{\circ}, b= 63^{\circ}$) are aligned with each other and with the 
direction of the dipole ($l=264^{\circ}, b= 48^{\circ}$). This indicates a preferred 
axis; and is inconsistent with the Gaussian random, statistically isotropic skies 
at about $99\%$.
\item Using HST key project data \cite{mclure07} showed that a statistically significant 
variation of at least 9 km/s/Mpc exists in the observed value of 
$H_0$ with a directional uncertainty of about 10-20\%. \cite{gup11} also found 
directional dependence in the HST key project data.
\item Recent work provides some evidence for what is known as the Hubble Bubble 
\citep{zeh98,jha07} which suggests that we might be living inside a large void. 
Value of the Hubble constant inside the bubble is different from what is outside the
bubble. There is evidence for such large scale voids in the CMB maps
as well \citep{cruz05,olv06,cruz06,cruz07}, suggesting that such large voids
are not implausible. If our position is near the center of such a void one can 
explain the dimming of SNe without invoking dark energy. However, this challenges
the Copernican principle that we do not occupy a special place in the Universe. 
If we assume that our position is off-centered then one can explain the preferred
axis (or tiny departures from CP) also (\cite{blom09}). 
\end{itemize}
In summary, some cosmological observations show departure from CP and hence are 
in contradiction to the standard model of cosmology. 
Another threat to the model comes from the fact that 
Observed value of $\Lambda$ does not match its theoretical value and requires
fine tuning. To avoid this 
several alternative explanations have been suggested \citep{silv09,frie08,sahn06}. 
In many of these alternative models, $w$ is allowed to be different from $-1$ in the past; 
and approaches $w \simeq -1$ at low red-shifts. It is different from cosmological 
constant, since there the value of $w$ is always $-1$. Cosmological observations 
suggest $w \simeq -1$ at the present epoch, which is consistent 
with $\Lambda$CDM as well as the alternative cosmologies. To be able to distinguish 
among various models, we require data that is precise enough to
discern tiny variations in the dark energy. It is also required that
data be available at a large number of redshifts to
constrain the detailed behavior of dark energy with time. At the
present the only data that comes reasonably close to these
requirements is provided by the observations of the high-redshift
supernovae, which are believed to be standard candles. Besides, the quality of the 
SNe Ia data maybe doubtful due to the following reasons:
\begin{itemize}
\item The physics of SNe Ia is relatively poorly understood.
\item The possibility of physical mechanisms, such as dust in the
inter-galactic medium that systematically dims them. 
\item The supernova data are usually collated from several different
sources that might have slightly different systematics, due either to
instrumental effects or to the fact that they occur in different
directions in the sky. Since we have to correct for the galactic
dust extinction, which might not get completely removed from the
samples, this might produce anisotropy in data. 
\end{itemize}
These considerations imply 
that to have precise information about the behavior of dark energy we 
should have a good knowledge of the statistical properties of supernovae, 
both random as well as systematic. There have been several attempts to 
search for the direction dependence in the SNe Ia data. 
\cite{gup08} (hereafter GSL08) and \citep{gup10} (hereafter GS10) 
used the extreme value statistics to show that the two supernova data sets,
\cite{rie04} (GD04) and \cite{rie07} (GD07), do show some evidence for
direction dependence. \cite{anton10} have also shown a preferred axis using
the Union2 catalogue from \cite{aman10}. Several other works have also indicated
either systematic problems with the high-redshift supernova data or
directional dependence in the supernova data and other probes
\citep{ness04,ness05,ness07, jain06,jain07,luca13}. 

In this paper our main task is 1) to look for direction dependent systematic effects 
in the latest SNe Ia data (Union 2 catalogue) and 2) to compare the quality of this 
data with the previous data sets (Gold data 2004 and 2007). 
The plan of the paper is as follows. In \S~2 we introduce the statistic we have used, 
in \S~3 we provide our results and end with conclusions in \S~4.

%***************************************
\section{Methodology: The $\Delta_{\chi^2}$ statistic}
\label{sec:method}
We use the Gold data (GD04 and GD07) \citep{rie04,rie07} and the Union2 data 
\cite{aman10} for our analysis. The Gold data GD04 and GD07 contain 157 and 
182 SNe respectively; while the most recent and largest set Union2 contains 557 Supernovae. 
For a given supernova the measured quantity, the distance modulus
$\mu$, is the difference between the apparent and the absolute
magnitude
\begin{equation}
 \mu(z) = m(z) - M\,, 
\label{eq:mu1}
\end{equation}
where the apparent magnitude $m(z)$ depends on the intrinsic
luminosity of a supernova, the redshift $z$ and the cosmological
parameters; and $M$ is the absolute magnitude of a type~Ia supernova. 
It can be expressed in terms of the luminosity distance
$D_L$ as
\begin{equation}
\mu(z) = 5 \log \left({D_L(z)/{\rm Mpc}} \right) + 25\,,
\label{eq:mu}
\end{equation}
where the luminosity distance is given by
\begin{equation}
D_L(z) = \frac{c (1+z)}{H_0}\int_{0}^{z} \frac{dx}{h(x)}\,,
\label{eq:D_L}
\end{equation}
where $h(z; \Om,\Ox)=H(z; \Om,\Ox)/H_0$, and thus depends only on the
cosmological parameters matter density $\Om$ and the dark energy
density $\Ox$. We assume that the prescription for the variation of
dark energy with redshift is separately specified, for example in the
$\Lambda$CDM model the energy density in the dark energy remains a
constant. In Eq~\ref{eq:mu1} the dependence of the measured quantity
$\mu$ on $M$ is linear. Since $\mu$ depends on the logarithm of the
luminosity distance it is clear that it depends linearly on the logarithm of
the Hubble parameter $H_0$. Usually the data is given in terms of
Eq~\ref{eq:mu}, where the constant
$M$ has already been marginalized over. Thus, instead of two
nuisance parameters we are left with only one parameter, the Hubble
constant $H_0$. 

%***********************
We now give an introduction of the $\Delta$ statistic previously
introduced and used in GSL08 and GS10. A generalization was given in GS10, where 
we had numerically marginalized over the Hubble constant $H_0$. 
Some of the details are repeated here to make
this work self-contained. For our analysis we have assumed a flat $\Lambda$CDM
universe. A similar analysis could be carried out for a more general
model of dark energy.

We consider subsets of the full data set to construct our statistic
consisting of $N_{\rm subset}$ data points. Since the $\Lambda$CDM model fits the 
gold data sets GD04, GD07 and Union2 well, we first obtain the best 
fit to the \emph{full data sets} by minimizing the $\chi^2$, which we define as:
\begin{equation}
\chi^2 = \Sigma_{i=1}^N {\big[ \frac{\mu^i - \mu^{\Lambda CDM}}{\sigma_i} \big]}^2 \, ,
\end{equation} 
where $\sigma_i$ is observed standard error in $\mu_i$. By this we obtain the 
best fit values of the parameters $\Omega_M$ and $H_0$. 
Then for each supernova we calculate \mbox{ $\chi_i = [\mu^i -
\mu^{\Lambda\rm CDM} (z_i;\Om)]/\sigma_i $}, where  $\mu^{\Lambda\rm CDM} (z_i;\Om)$ 
is calculated using the best fit values of $\Omega_M$ and $H_0$. 
We assume that all the supernovae are statistically uncorrelated.

We define $\chi^2_M = \Sigma_i \chi_i^2 $ and the normalized quantity $\chi^2_R = \chi^2_M/N_{\rm subset}$,
$\chi^2_R$ indicates the statistical scatter of the subset from the best fit $\Lambda$CDM model
and its expectation value is unity that is
$\langle \chi^2_R \rangle = 1$.  If CP holds then the apparent
magnitude of a supernova should not depend on the direction in which
it is observed but only on the cosmology and thus supernovae in
different directions should be scattered similarly with respect to the
best fit model.
%******************
\begin{figure}
\centering
\includegraphics[width=0.45\textwidth]{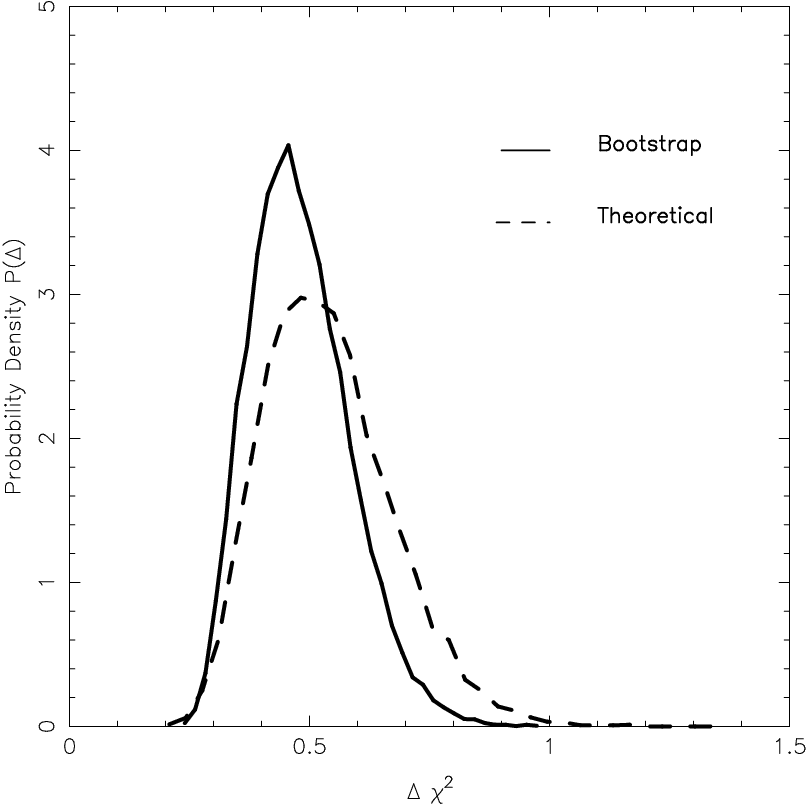}
\caption{A comparison of theoretical and bootstrap probability distributions for 
simulated data. The data comprises $157$ supernovae, whose positions on the sky 
were generated randomly.}
\label{fig:sim157}
\end{figure}

\begin{figure}
\centering
\includegraphics[width=0.45\textwidth]{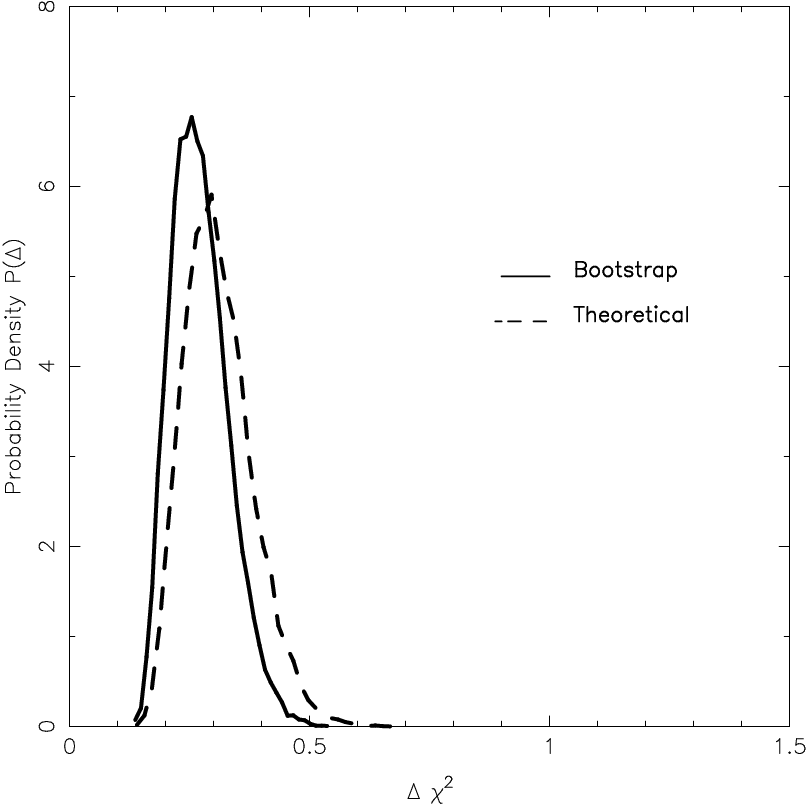}
\caption{A comparison of theoretical and bootstrap probability distributions for simulated data. 
The data comprises $557$ supernovae, whose positions on the sky were generated randomly.}
\label{fig:sim500}
\end{figure}
%**********************

We divide the data into two hemispheres labeled by the direction
vector $\hat{n}$, and take the difference of the $\chi_R^2$
computed for the two hemispheres separately to obtain $\Delta
\chi_{\hat{n}}^2 =\chi^2_{R1} - \chi^2_{R2} $, where label '1' corresponds to
that hemisphere towards which the direction
vector $\hat{n}$ points and label '2' refers to the other hemisphere. 
We take the absolute value of $\Delta
\chi_{\hat{n_i}}^2$ since we are interested in the largest value of this quantity
and it is obvious that for every value of $\Delta
\chi_{\hat{n_i}}^2$, the antipodal point has the negative of that value. 
We then vary the direction
$\hat{n}$ across the sky to obtain the maximum absolute difference
\begin{equation}
\Delta_{\chi^2} = {\rm max} \{| \Delta \chi_{\hat{n}}^2 |\}\,\,. 
\end{equation}
As shown in GSL08, the distribution of $\Delta_{\chi^2}$ follows a
simple, two parameter Gumbel distribution, characteristic of extreme
value distribution type~I \citep{ken77},
\begin{equation}
P(\Delta_{\chi^2}) =\frac{1}{s} \exp \left[ -\frac{ \Delta_{\chi^2}
-m}{s}\right]\,\exp\left [-\exp\left(-\frac{\Delta_{\chi^2}
-m}{s}\right)\right]\,\,,
\end{equation} 
where the position parameter $m$ and the scale parameter $s$
completely determine the distribution. To quantify departures from
isotropy we need to know the theoretical distribution,
which is calculated numerically by simulating several
sets of Gaussian distributed $\chi_i$ on the gold set supernova
positions and obtaining $\Delta_{\chi^2}$ from each realization. And
as in GSL08, we compute a bootstrap distribution by shuffling the data 
values $z_i$, $\mu(z_i)$ and $\sigma_{\mu}(z_i)$ over
the supernovae positions (for further details see GSL08).
A modified version of this statistic was introduced in GS10, 
which marginalizes over the Hubble Constant. However, by this we 
loose all the information about $H_0$ and hence, here we do 
not marginalize. 

%******************************
\section{Results}
\begin{figure}
\centering
\includegraphics[width=0.45\textwidth]{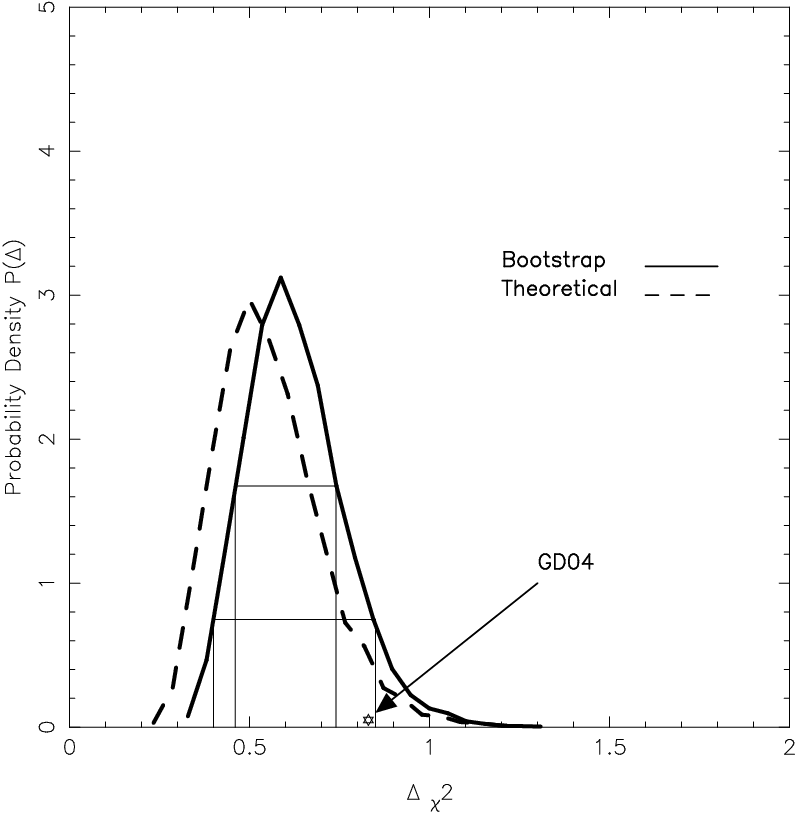}
\caption{The theoretical and the bootstrap probability distributions for GD04 for the $\Delta_{\chi^2}$ statistic. Theoretical
distribution is shifted to the left compared to what we find for our simulated data
in Figure~\ref{fig:sim157}, which uses Gaussian deviates suggesting evidence for non-Gaussianity.}
\label{fig:old04}
\end{figure}
%******************************
\begin{figure}
\centering
\includegraphics[width=0.45\textwidth]{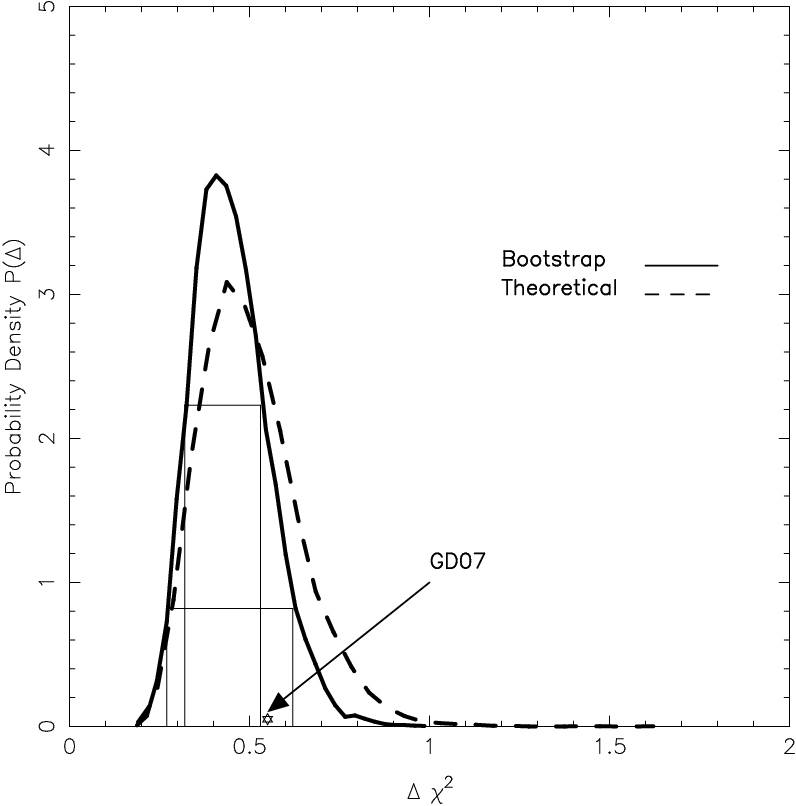}
\caption{The theoretical and the bootstrap probability distributions for GD07 for the $\Delta_{\chi^2}$ statistic. Theoretical
distribution is to the right as expected by simulated data in Figure~\ref{fig:sim157}, 
which uses Gaussian deviates suggesting evidence for non-Gaussianity.}
\label{fig:old07}
\end{figure}
%******************************
\begin{figure}
\centering
\includegraphics[width=0.45\textwidth]{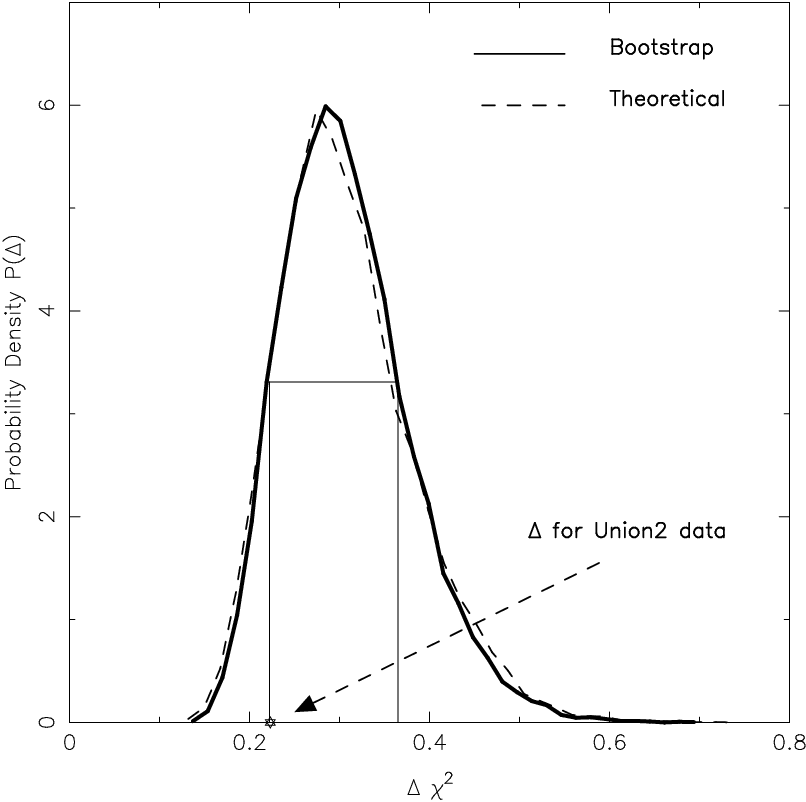}
\caption{The theoretical and the bootstrap probability distributions for Union2 for the $\Delta_{\chi^2}$ statistic. 
Theoretical distribution is not compatible with Figure~\ref{fig:sim500} for the simulated data, which 
implies slight non-Gaussianity for the residuals.}
\label{fig:union2chisq}
\end{figure}
%******************************
In GSL08 and GS10 we discussed a specific bias in the bootstrap distribution,
showing that it is shifted slightly to the left of the theoretical
distribution due to the fact that theoretical distribution is obtained
by assuming $\chi_i$s to be Gaussian random variates with a zero mean
and unit variance. Therefore theoretical $\chi_i$s are
unbounded. However, the bootstrap distribution is obtained by
shuffling through a \emph{specific realization} of $\chi_i$, and they
have a maximum value for some supernova. It is clear that this should,
on the average, produce slightly smaller values of $\Delta_{\chi^2}$
in comparison with what one expects from a Gaussian distributed
$\chi_i$. For reference we plot the results for simulations 
in Figure~\ref{fig:sim157} with a total of $157$ and in Figure~\ref{fig:sim500} 
with a total of $557$ supernovae. Our results in this paper should be interpreted with 
respect to Figure~\ref{fig:sim157} for GD04 and GD07, and Figure~\ref{fig:sim500} for Union2 data.
Concerns regarding the small number of
supernovae in the Gold data and its effect on the efficacy of our
method can bee addressed by the fact that overall behavior of Figure~\ref{fig:sim157} is 
repeated in Figure~\ref{fig:sim500} which is plotted for more than $500$ of SNe.

%****************
\begin{table}
\begin{center}
\caption{
  The model parameters ($\Lambda$CDM) for the three data sets are tabulated  here.}  
  \label{tbl-oldfit}
   \bigskip

\begin{tabular}{ccrrrrr}
\hline
 Set & \# SNe & $\Om$ &  $H_0$  &  $\chi^2/dof$ \\
\hline
GD04  &  157 & 0.32 & 64.5 &  1.14 \\
GD07  &  182 & 0.33 & 63.0 &  0.88 \\
Union2 & 557 & 0.27 & 70  &  0.97 \\
\hline\hline                                                                                        
\end{tabular}
\end{center}
\end{table}
%****************
\begin{table}
\begin{center}
\caption{
  Direction for maximum $\Delta$ in the three data sets are tabulated  here.}
  \label{tbl-olddel} \bigskip

\begin{tabular}{crrrrrr}
\hline
Model & Set & $\Delta_{\chi^2}$ &  longt  &  lat \\
\hline
$\Lambda$CDM & GD04 & 0.83 & 96.5 &  44.5 \\
$\Lambda$CDM & GD07 & 0.53 & 347.1 &  27. \\
$\Lambda$CDM & Union2 & 0.22 & 65.5 &  55.8 \\
\hline\hline
\end{tabular}
\end{center}
\end{table}
%*******************************
This statistic is similar to the one presented in GSL08 and GS10 except 
for the fact that in GS10 we had marginalized over the Hubble parameter. Our
results are different from those presented in GSL08 also due to the
fact that we have corrected the numerical bug mentioned in the
GS10 produced by the fact that the theoretical distribution
was produced slightly differently from the way bootstrap method thereby 
creating a greater discrepancy between them then should have been the case.
In Table~\ref{tbl-oldfit} we give the best fit values of $\Om$ and $H_0$ 
for all the three data sets. We note that Union2 gives slightly lower value of $\Om$ 
and higher value of $H_0$ compared to the Gold data. Reduced $\chi^2$ is smallest for 
GD07 which indicates that errors are overestimated for this set. Union2 is best 
among all the three sets in terms of $\chi^2$. The direction of maximum discrepancy
and the value of ($\Delta_{\chi^2}$) is presented in Table~\ref{tbl-olddel}. 

{\bf GD04}: In Fig~\ref{fig:old04} we plot the bootstrap and the theoretical distribution expected 
for GD04 and mark the position of GD04. Comparison with Figure~\ref{fig:sim157} 
shows a signature of non-Gaussianity since the theoretical 
distribution instead of being to the right of the bootstrap is shifted to the left. Position 
of GD04 is about 2 sigma away from the peak of the bootstrap distribution, indicating 
direction dependence. 

{\bf GD07}: Fig~\ref{fig:old07} for GD07 is same as for GD04. 
A comparison with Figure~\ref{fig:sim157} shows that our
results are compatible with the absence of non-Gaussianity
in the data. GD07 sits at about one sigma away from the peak of the distribution, 
thus the directional dependence has got weaker than found in GD04.

{\bf Union2} In Fig~\ref{fig:union2chisq} we plot both the distributions for Union 2,
which is to be compared with Fig~\ref{fig:sim500}. Comparison shows a weak signature of 
non-Gaussianity, since position of the bootstrap distribution is not to the left of the 
theoretical distribution. Position of Union2 is about one sigma away from the peak 
of the bootstrap distribution which is a sign of direction dependence.

%********************************
\section{Conclusions}
We have presented results for the GD04, GD07 and Union2 data using the statistic 
introduced in GSL08 and GS10. Our main conclusions for this part of our work are:
\begin{enumerate}
\item GD04 shows some evidence for non-Gaussianity, however, GD07 is entirely consistent with
a Gaussian distribution of residuals. Union 2 data again shows some evidence for non-Gaussianity
although weaker compared to GD04. Thus, in terms of Gaussianity, GD07 is best among the three
data sets.
\item GD04 is different from the peak of the bootstrap distribution by slightly more than one 
sigma; while GD07 and Union2 both are different from the peak by about a sigma.
There seems a weak but consistent direction dependence in all three data sets. 
This consistency indicates a physical effect or a preferred direction.
\end{enumerate}
%************************************
\medskip
\noindent{\it Acknowledgments:}

Shashikant Gupta thanks Tarun Deep Saini for discussion; and colleagues 
of ASAS for constant support. 

%************************************

\def\etal{{\it et~al.\ }}
\def\apj{{Astroph.\@ J.\ }}
\def\araa{{Ann. \@ Rev. \@ Astron. \@ Astroph.\ }}
\def\mn{{Mon.\@ Not.\@ Roy.\@ Ast.\@ Soc.\ }}
\def\asta{{Astron.\@ Astrophys.\ }}
\def\aj{{Astron.\@ J.\ }}
\def\prl{{Phys.\@ Rev.\@ Lett.\ }}
\def\pd{{Phys.\@ Rev.\@ D\ }}
\def\nucp{{Nucl.\@ Phys.\ }}
\def\nat{{Nature\ }}
\def\sci{{Science\ }}
\def\plb {{Phys.\@ Lett.\@ B\ }}
\def \jetpl {JETP Lett.\ }
\def \jcap {J. Cosmol. Astropart. Phys.}

\end{document}